\documentstyle[emulateapj,onecolfloat]{article}

\newcommand{\bn}{\hat{\bf n}}
\newcommand{\be}{{\bf e}}
\newcommand{\bfl}{{\bf l}}
\newcommand{\ApJL}{ApJ Lett.}
\newcommand{\ApJ}{ApJ}

\newcommand{\PRD}{Phys. Rev. D}
\newcommand{\MNRAS}{MNRAS}
\newcommand{\ARAA}{ARA\&A}
\newcommand{\AsAs}{A\&A}

\newcommand{\aut}[2]{{#1,\ #2.,}}
\newcommand{\laut}[2]{\& {#1,\ #2.}}
\newcommand{\refs}[6]{#5, #2, #3  {#4}.}
\newcommand{\urefs}[5]{#4, #2, #3 (#5).}
\newcommand{\mybib}[2]{\bibitem[#1]{#2}}
\input epsf

\begin{document}
\twocolumn[
\title{Power Spectra Estimation for Weak Lensing}
\author{Wayne Hu$^1$ and Martin White$^{2}$}
\affil{{}$^1$Department of Astronomy and Astrophysics,
		University of Chicago, Chicago IL 60637 \\
       {}$^2$Harvard-Smithsonian CfA, 60 Garden St, Cambridge, MA 02138}
\begin{abstract}
We develop a method for estimating the shear power spectra from weak lensing
observations and test it on simulated data.  Our method describes the shear
field in terms of angular power spectra and cross correlation of
the two shear modes which differ under parity transformations.  Two
of the three power spectra can be used to monitor unknown sources
of noise in the data.  The power spectra are decomposed in a model
independent manner in terms of ``band-powers'' which are then 
extracted from the data using
a quadratic estimator to find the maximum  of the likelihood and its local
curvature (for error estimates).  We test the method against simulated
data from Gaussian realizations and cosmological $N$-body simulations.
In the Gaussian case, the mean bandpowers and their covariance are
well recovered even for irregular (or sparsely) sampled fields.  The
mild non-Gaussianity of the $N$-body realizations causes a slight 
underestimation of the errors that becomes negligible for scales
much larger than several arcminutes and does not bias the recovered
band powers.
\end{abstract}
]
\section{Introduction}

Weak lensing of background galaxies by foreground large-scale structure
has now been convincingly detected
(\cite{BacRefEll00} 2000; \cite{KaiWilLup00} 2000; \cite{vanetal00} 2000;
\cite{Witetal00} 2000)
and is rapidly becoming a valuable tool for studying the distribution and
clustering evolution of dark matter in the universe.
Lensing produces a correlated distortion of the ellipticities of background
galaxies, at the percent level, which can be used to measure a two-dimensional
projection of the mass distribution. 

While there are numerous observables that can be defined from the shear
maps produced by such surveys, one of the most important is the angular
power spectrum.  The angular power spectrum contains valuable information
on cosmological parameters that complements other astrophysical
measurements (\cite{JaiSel97} 1997; \cite{Kai98} 1998; \cite{HuTeg99} 1999).
In this respect weak lensing is very similar to anisotropies
in the CMB and in fact much of the theory and data analysis
is very analogous as well.

In this paper, we suggest a technique for the presentation and interpretation
of weak lensing data which is commonly employed in CMB analysis:
the use of ``band-powers'' extracted from the data by an iterated quadratic
estimator of the maximum likelihood solution.
This technique has the advantage of automatically taking into account
irregular survey geometries and varying sampling densities.  It provides an
optimal estimate of the power spectrum in the Gaussian regime and
makes efficient use of all of the data on the relevant angular scales.
Error estimates that include the sampling and noise variance of the
survey are also automatically produced by this method.
Similar claims cannot be made for methods based on correlation functions
or simple Fourier Transforms of the data.

Throughout our focus will be on weak lensing by large-scale structure,
specifically on angular scales larger than several arcminutes (or
wavenumbers $l \la 10^3$).
These are the scales which will be probed by ongoing wide field surveys,
e.g.~the Deep Lens Survey\footnote{http://dls.bell-labs.com}, surveys
with the VLT and the Hawaii/IfA lens survey.
On these angular scales the signal reduces simply to a projection of the
density contrast along the line of sight
(\cite{BlaSauBraVil91} 1991; \cite{Mir91} 1991; \cite{Kai92} 1992), 
which is well approximated
by a Gaussian probability distribution.  In this limit the angular power
spectrum encodes all of the relevant information about the field, and in
particular can be predicted from the 3D power spectrum of the density field
using Limber's equation.
On subarcminute scales, the field becomes substantially 
non-Gaussian, lens-lens coupling and perturbations to the photon 
trajectory become
increasingly important and the weak lensing approximations break down.
On these angular scales the signal is dominated by individual objects along
the line of sight (e.g.~clusters of galaxies) and a correlation function or
power spectrum analysis becomes less useful.

The outline of this paper is as follows.  In \S \ref{sec:theory} we
describe the shear signal covariance matrix and its relationship to
the three fundamental shear power spectra.
We then develop  (\S\ref{sec:likelihood}) and test (\S\ref{sec:tests})
the iterated likelihood method for band power extraction using simulated
data.  We conclude in \S\ref{sec:conclusions}.

\begin{figure*}[tbh]
\centerline{\epsfxsize=6truein\epsffile{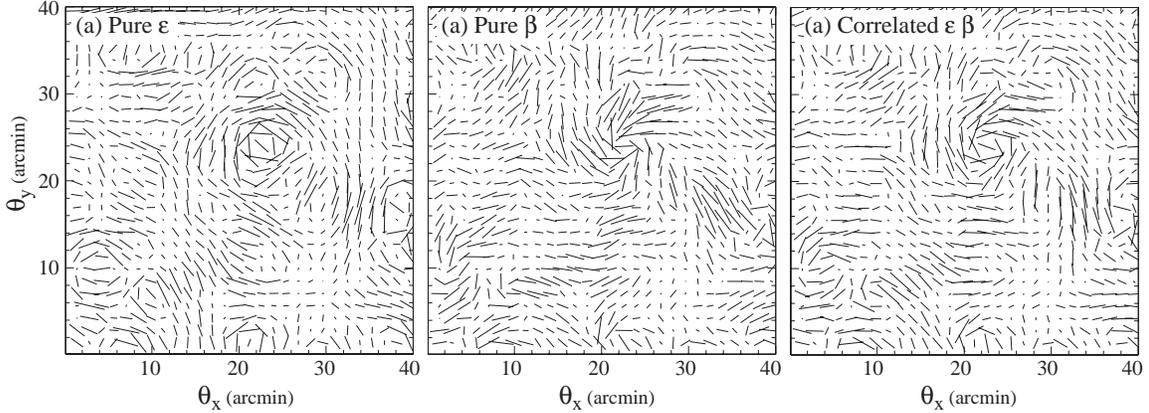}}
\caption{\footnotesize The fundamental shear modes and their
cross correlation.  
(a) A pure $\epsilon$-field obtained from a convergence
map from \protect{\cite{WhiHu00}} (2000).  (b) A pure $\beta$-field obtained by a rotation of the shears by $\pi/4$.  (c) A correlated mixture of $\epsilon$ and $\beta$ 
with $C_l^{\epsilon\epsilon}=C_l^{\beta\beta}=C_l^{\epsilon\beta}$ 
obtained by a rotation of the shears by $\pi/8$.}
\label{fig:shear}
\end{figure*}

\section{Weak Lensing} \label{sec:theory}

In this section we briefly review the theory of weak gravitational lensing
to establish our notation and conventions.  Throughout we shall focus on the
2-point function of gravitational shear, since this shall determine the
angular power spectrum in \S\ref{sec:likelihood}.

The gravitational deflection of light induces a mapping between the 2D source
plane (S) and the image plane (I).  The deformation so induced can be written
\begin{equation}
  \delta x^{\rm S}_i = A_{ij} \delta x^{\rm I}_j\,,
\end{equation}
where $\delta {\bf x}$ is the separation vector between points on the
respective planes.  In the weak lensing limit, the deformation
can be decomposed as 
(\cite{Mel99} 1999;
\cite{BarSch00} 2000)
\begin{equation}
  A_{ij} = (1-\kappa)\delta_{ij} - \gamma_1 \sigma_3 - \gamma_2 \sigma_1 \,,
\end{equation}
where the $\sigma_i$ are the $2\times 2$ Pauli matrices, $\kappa\ll1$ is the
convergence and $\gamma_a\ll1$ is the shear.  If a galaxy has (weighted)
second moments $M^{S}$ then the image will have
\begin{equation}
  M^{I} = A^{-1}\cdot M^{S}\cdot A^{-1}\,.
\label{eqn:secondmoments}
\end{equation}
The ellipticities are usually defined in terms of the second moments of the
light distribution, corrected for instrumental and observational effects,
and in the weak lensing regime, Eq.~(\ref{eqn:secondmoments}) simplifies
dramatically such that the observed ellipticity of a galaxy is 
linearly related to the shear.  
The proportionality constant depends on the definition of
the ellipticity; we take
\begin{eqnarray}
 \left< \be \right> = \hbox{\boldmath $\gamma$ \unboldmath}\,,
\end{eqnarray}
but note that 2\hbox{\boldmath $\gamma$
\unboldmath} is often found in the literature
(\cite{BarSch00} 2000).
The result is that  $\be$ defines a
(noisy) estimate of the local shear field at $\bn$.

Now consider an observation of a given area of the sky.
The observed field yields an estimate of the ellipticities $\be_i$ and
positions $\bn_i$ of a set of galaxies binned into pixels 
$i=1,\cdots N_{\rm pix}$.
In a Cartesian coordinate system on the sky
the two components of the shear field, $\gamma_1(\bn)$ and $\gamma_2(\bn)$,
transform as a spin-2 field.  The Fourier decomposition is
\begin{equation}
  \gamma_1(\bn) \pm i \gamma_2(\bn) = 
  \int {d^2 l \over (2\pi)^2} W(\bfl) 
  [\epsilon (\bfl) \pm i \beta(\bfl)]
  e^{\pm 2 i \varphi_l} e^{i \bfl \cdot \bn} \,,
\end{equation}
where $\varphi_l$ is the angle between $\bfl$ and the x-axis and $W(\bfl)$
is the Fourier transform of the pixel window function.  For square pixels
of side $\sigma$ in radians
\begin{equation}
W(\bfl) = 
j_0\left( {l \sigma \over 2} \cos\varphi_l \right)
j_0\left( {l \sigma \over 2} \sin\varphi_l \right) \,,
\end{equation} 
where $j_0(x)=\sin(x)/x$ is the 0th order spherical Bessel function.
Note that for long wavelengths the pixelization is irrelevant and the
window goes to unity.

We shall be interested in the power spectrum or correlation function of the
shear field.  The two point correlations in the shear are determined by 
the three shear power spectra
\begin{eqnarray}
 \left< \epsilon(\bfl) \epsilon(\bfl') \right> &=& (2\pi)^2 \delta(\bfl-\bfl') 
 C_l^{\epsilon\epsilon}\,, \nonumber\\
 \left< \beta(\bfl) \beta(\bfl') \right> &=& (2\pi)^2 \delta(\bfl-\bfl') 
 C_l^{\beta\beta}\,, \nonumber\\
 \left< \epsilon(\bfl) \beta(\bfl') \right> &=& (2\pi)^2 \delta(\bfl-\bfl') 
 C_l^{\epsilon\beta}\,, 
\end{eqnarray}
For the shear generated by weak lensing
$C_l^{\epsilon\epsilon} = C_l^{\kappa\kappa}$,
$C_l^{\beta\beta}=0$ and $C_l^{\epsilon\beta}=0$.  
For shot noise $C_l^{\epsilon\epsilon}=C_l^{\beta\beta}$.
Systematic errors can in principle generate any of the power spectra.

Since a $45^\circ$ degree rotation of the shears
takes $\epsilon \rightarrow \beta$, it converts the lensing signal to
a $C_l^{\beta\beta}=C_l^{\kappa\kappa}$ spectrum with $C_l^{\epsilon\epsilon}
=C_l^{\epsilon\beta}=0$.
A more general rotation leaves a signal in both $C_l^{\epsilon\epsilon}$
and $C_l^{\beta\beta}$ but also correlates them as $(C_l^{\epsilon\beta})^2
= C_l^{\epsilon\epsilon}C_l^{\beta\beta}$.  
For the shot noise, the relation 
$C_l^{\epsilon\epsilon}= C_l^{\beta\beta}$ 
is invariant under rotations. 
These rotations
also allow one to visualize the pattern implied by each spectra 
(see Fig.~\ref{fig:shear}).
In particular, the $\beta$ component possesses a ``handedness''; formally
the two are distinguished by their transformation under parity.

By direct substitution
\begin{eqnarray}
\left< \gamma_1(\bn_i) \gamma_1(\bn_j) \right> 
	&=&
	\int {d^2 l \over (2\pi)^2} 
		\Big(C_l^{\epsilon\epsilon} \cos^2 2\varphi_l
		+      C_l^{\beta\beta} \sin^2 2\varphi_l 
	\nonumber \\&&\quad 
		-C_l^{\epsilon\beta}\sin 4\varphi_l \Big) W^2(\bfl)
		e^{i \bfl \cdot (\bn_i-\bn_j)}\,,
	\nonumber\\
\left< \gamma_2(\bn_i) \gamma_2(\bn_j) \right> 
	&=&
	\int {d^2 l \over (2\pi)^2} 
		\Big( C_l^{\epsilon\epsilon} \sin^2 2\varphi_l
		+      C_l^{\beta\beta} \cos^2 2\varphi_l 
	\nonumber\\ &&\quad 
		+C_l^{\epsilon\beta}\sin 4\varphi_l \Big) W^2(\bfl)
		e^{i \bfl \cdot (\bn_i-\bn_j)}\,,
	\nonumber\\
\left< \gamma_1(\bn_i) \gamma_2(\bn_j) \right> 
	&=&
	\int {d^2 l \over (2\pi)^2} 
		\Big[ {1 \over 2} ( C_l^{\epsilon\epsilon}
		-          C_l^{\beta\beta} ) \sin 4\varphi_l 
	\nonumber\\ &&\quad 
		+	   C_l^{\epsilon\beta} \cos 4\varphi_l \Big] W^2(\bfl) 
		e^{i \bfl \cdot (\bn_i-\bn_j)} \,.
\label{eqn:ggdef}
\end{eqnarray}
For a coordinate system that is oriented so that ${\bn_i-\bn_j} \parallel
{\bf x}$ and pixel separations that are small compared with the 
coherence scale of the field, the cosmological $\epsilon\epsilon$ signal 
generates $\left< \gamma_1 \gamma_1 \right> > 0$.
For shot noise $\left< \gamma_2 \gamma_2 \right>
\approx  \left<\gamma_1 \gamma_1 \right>$ and for either 
$\left< \gamma_1 \gamma_2 \right>\approx 0$.  These are the tests suggested
by \cite{Mir91} (1991). 

\begin{figure}[tbh]
\centerline{\epsfxsize=3.25truein\epsffile{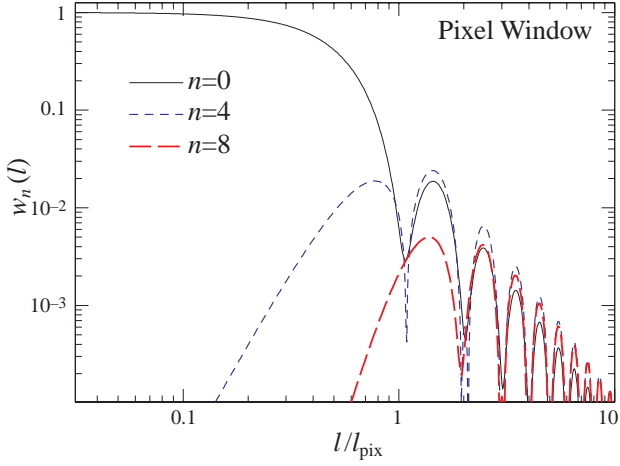}}
\caption{\footnotesize Low order moments of the pixel window function squared. The first
two moments sufficiently approximate the pixel window function to
$l \sim 2 l_{\rm pix} = 4\pi/\sigma$ where $\sigma$ is the width of
the pixel in radians.}
\label{fig:window}
\end{figure}

These relations (\ref{eqn:ggdef}) allow us to define the lensing signal correlation matrix
\begin{equation}
  C_{(ij)(ab)}^{\rm sig} = \left< \gamma_a(\bn_i) \gamma_b(\bn_j) \right>\,.
  \label{eqn:cs}
\end{equation} 
The correlation matrix may be simplified by recalling that
\begin{equation}
  \exp\left( i \bfl\cdot{\hbox{\boldmath{$\theta$}\unboldmath}} \right) =
    J_0(l \theta) + 2\sum_{m=1}^{\infty} i^m J_m(l\theta) 
    \cos(m(\varphi_l - \phi))\,,
\end{equation}
where $(\theta,\phi)$ define the magnitude and orientation of the
separation vector $\bn_i - \bn_j$.  Furthermore, the window function
can be similarly decomposed as
\begin{equation}
W^2(\bfl) = \sum_{n=0}^\infty w_n(l) \cos(n \varphi_l) \,.
\end{equation}
For square pixels, the $n=1,2,3$ moments vanish due to symmetry 
and it is sufficient to retain the isotropic and $n=4$ quadrupole 
contributions (see Fig.~\ref{fig:window}).  The 
integral over $\varphi_l$ can now be performed analytically
leaving
\begin{equation}
  C_{(ij)(ab)}^{\rm sig} = \int {l d l\over 4\pi}
    \sum_X C_l^{X} \left[w_0(l) I_{ab}^{X} + {1 \over 2} w_4(l) 
    Q_{ab}^{X}\right]\,,
\label{eqn:signalcov}
\end{equation}
where $X$ takes on the values $\epsilon \epsilon$, $\beta \beta$ and
$\epsilon \beta$,
\begin{eqnarray}
  I^{\epsilon\epsilon} &=& \left[
  \begin{array}{cc} J_0+c_4 J_4 &      s_4 J_4 \\
                        s_4 J_4 & J_0 - c_4 J_4 \\ \end{array}
  \right]\,,\nonumber\\
  I^{\beta\beta} &=& \left[
  \begin{array}{cc} J_0-c_4 J_4 &     -s_4 J_4 \\
                       -s_4 J_4 & J_0 + c_4 J_4 \\ \end{array}
  \right]\,,\nonumber\\
  I^{\epsilon\beta}&=& \left[
  \begin{array}{cc}   -2 s_4 J_4 &      2 c_4 J_4 \\
            \phantom{-}2 c_4 J_4 &      2 s_4 J_4 \\ \end{array}
  \right]\,,
\end{eqnarray}
and
\begin{eqnarray}
  Q^{\epsilon\epsilon} & = &\left[
  \begin{array}{cc} J_0+ 2 c_4 J_4 + c_8 J_8 &      s_8 J_8 \\
                        s_8 J_8 & - J_0 + 2 c_4 J_4 - c_8 J_8\\ \end{array}
  \right]\,,\nonumber\\
  Q^{\beta\beta} & = &     \left[
  \begin{array}{cc} -J_0 + 2 c_4 J_4 - c_8 J_8 &     -  s_8 J_8 \\
                       - s_8 J_8 & J_0 + 2 c_4 J_4 - c_8 J_8 \\ \end{array}
  \right]\,,\nonumber\\
  Q^{\epsilon\beta} & = & \left[
  \begin{array}{cc}   -2 s_8 J_8 &      2 J_0 + 2c_8 J_8 \\
            \phantom{-}2 J_0 + 2 c_8 J_8 &      2 s_8 J_8\\ \end{array}
  \right]\,.
\end{eqnarray}
Here we have used the shorthand notation $c_n = \cos(n\phi)$ and
$s_n = \sin(n\phi)$ and the (suppressed) argument of the Bessel function
in each case is $l\theta$.
Thus for a flat bandpower we need only evaluate 
\begin{equation}
  C^{(n)}(\theta)\equiv \int_{l \in B} {dl\over 2l}
  \ J_n(l \theta)\,,
\label{eqn:corrn}
\end{equation}
for $n=0,4,8$.

\section{Shear Likelihood} \label{sec:likelihood}

We wish to estimate the (angular) power spectra of the shear field from
the observed image ellipticities by means of a maximum likelihood technique.
This ensures that, under the stated assumptions, we make optimal use of the
data and correctly handle any irregular survey geometry which may affect the
correlations on large angular scales.  Non-uniform or correlated noise can
also be efficiently handled by this formalism.

First we decide to parameterize the underlying power spectra with a set of
parameters $p_\alpha$ where $\alpha =1,\ldots,N_p$.  These parameters could be
cosmological, or describe a particular model of noise or systematic errors.
In this paper we shall mainly be interested in the case where the
$p_\alpha$ are ``bandpowers'',
i.e.~we shall approximate the angular power spectra as
piecewise constant with $p_\alpha$ the value of $l(l+1)C_l/2\pi$ in
band $B_\alpha$.  So long as the constant sections are narrower than any
feature in the power spectrum which we wish to reproduce, the sharp steps
in power will not produce any ill effects, and at the same time the
parameterization is model independent.

Such a specification is not sufficient to describe the most general likelihood
function of the observations given the $p_\alpha$, however if the shear field
is Gaussian it is.  We shall assume that on sufficiently large angular scales,
which are of interest here, the field is sufficiently Gaussian that the
estimate of the 2-point function so derived is not seriously in error.
Once the field becomes significantly non-Gaussian the utility of a power
spectrum estimate become suspect.
We shall return to this point in detail in the next section.

Consider the data as a $2 N$ component vector
\begin{equation}
  d = \{ \gamma_1(\bn_1),\gamma_2(\bn_1);\ldots;
         \gamma_1(\bn_N),\gamma_2(\bn_N)\}\,,
\end{equation}
then the likelihood is simply
\begin{equation}
  {\cal L}(p_\alpha) = 
  {1 \over (2\pi)^{N} |C(p_\alpha)|^{1/2}}
  \exp[ - {1 \over 2} d^T C^{-1}(p_\alpha) d] \,.
\end{equation}
Here the correlation matrix $C$ is a sum of two terms.  The first is the
cosmological signal, Eq.~(\ref{eqn:cs}), and the second is the noise.
Each galaxy has an rms intrinsic ellipticity per component 
${\bf \gamma}^{\rm int}$ which we assume is uncorrelated with the underlying
shear field taken to be constant across the galaxy.
The ellipticity is thus a noisy estimator of the shear field at its position,
and the noise matrix is
\begin{equation}
  C_{(ij)(ab)}^{\rm noise} ={\left( \gamma^{\rm int}_i \right)^2 \over N_i}
                             \delta_{ij} \delta_{ab}\,,
\end{equation}
where $N_i$ is the number of galaxies in pixel $i$.
In more generality, the noise matrix can include observational errors on the
ellipticities which could be correlated from galaxy to galaxy.  The likelihood
formalism can efficiently handle a general noise matrix.

One then maximizes the likelihood as a function of the model parameters
$p_\alpha$.
We accomplish this maximization iteratively by using
the Newton-Raphson method to find the root of $d{\cal L}/dp_\alpha=0$
(\cite{BonJafKno98} 1998; \cite{Sel98} 1998).
Specifically, from an initial set $\hat{p}_\alpha$ one makes an improved
estimate $p_\alpha = \hat{p}_\alpha + \delta p_\alpha$ where 
\begin{equation}
  \delta p_\alpha = \lambda \sum_\beta {1 \over 2} (F^{-1})_{\alpha \beta}
		  {\rm tr}\left[ (d d^T - \hat{C})(
			\hat{C}^{-1}
			\hat{C}_{,\beta}
			\hat{C}^{-1}) \right]\,,
\end{equation}
and where the Fisher matrix is
\begin{eqnarray}
F_{\alpha\beta} &=& {1 \over 2} {\rm tr} (\hat{C}^{-1} 
		\hat{C}_{,\alpha} \hat{C}^{-1} \hat{C}_{,\beta})\,,
\end{eqnarray} 
and $\lambda \le 1$ is a parameter that is set to reduce the step 
size when the full value causes large jumps in parameter space.
Note that in maximizing this function for the bandpowers several matrices,
and their derivatives, must be computed and inverted.
Using Eqs.~(\ref{eqn:signalcov}-\ref{eqn:corrn}),
we first compute the derivative matrices
$\hat{C}_{,\alpha}$ for $p_\alpha=1$.
The full correlation function is then the sum
$\sum_\alpha p_\alpha\hat{C}_{,\alpha}$ plus the noise term.

\begin{figure*}[tbh]
\centerline{\epsfxsize=5.5truein\epsffile{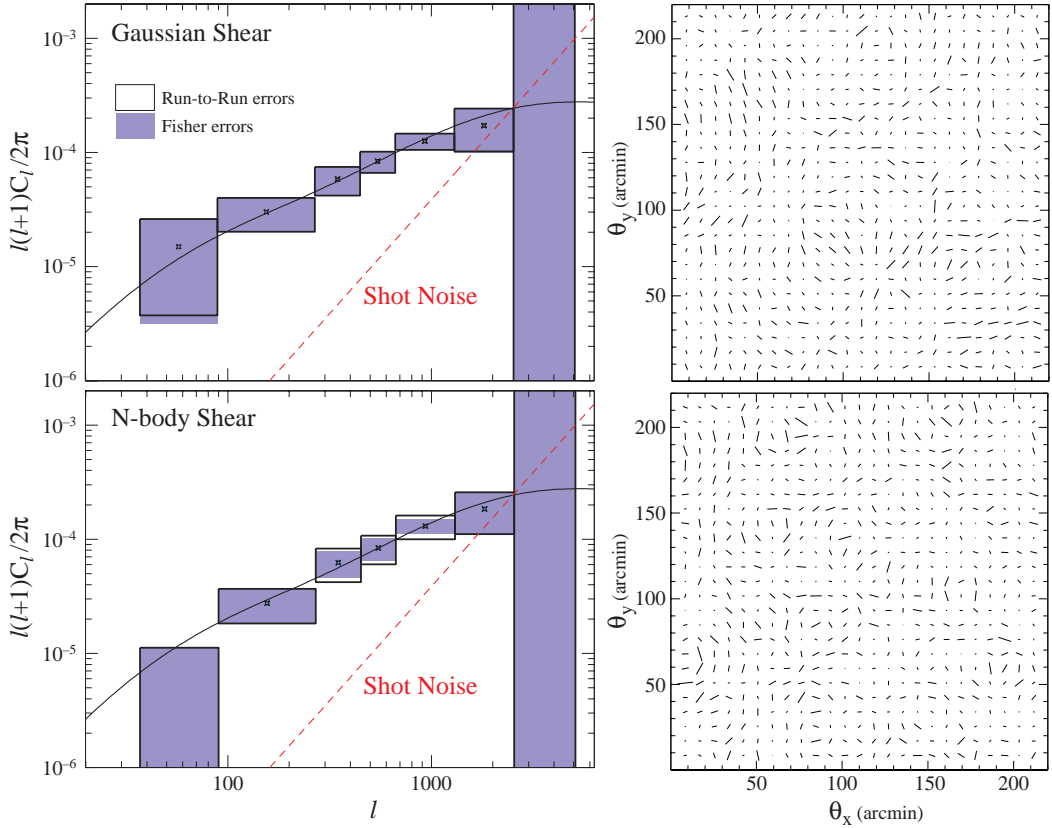}}
\caption{\footnotesize Gaussian (top) and N-body (bottom) 
simulated shear (right) and recovered $\epsilon\epsilon$ band powers (left)
from 200 realizations.  Solid lines represent the predicted power spectrum; 
dashed lines represent the assumed shot noise contribution.  Errors are 
estimated from the Fisher matrix (shaded boxes) and the run-to-run
scatter (open boxes). The simulated fields contain 625 pixels 
and are analyzed with 7 $\epsilon\epsilon$ bands, 6 $\beta\beta$ bands, 
and 6 $\epsilon\beta$ bands.  The latter two sets show no significant recovered 
power and have been omitted here for clarity.          
}
\label{fig:sim}
\end{figure*}

With an appropriate choice of $p_\alpha$ and $\lambda$
this method typically converges rapidly to the desired solution. 
General rules of thumb for these quantities include that the bands
should be at least twice as wide as $l_{\rm field}=2\pi/\theta_{\rm max}$ 
where $\theta_{\rm max}$ is the largest pixel separation.  This reduces
the covariance between the bands that is due to the survey geometry.
The lowest band should include $\epsilon\epsilon$ modes below $l_{\rm field}$
to absorb any d.c.~offsets in the data.  The highest band should
include modes above $l_{\rm pix} =2 \pi/\theta_{\rm pix}$ since the
window function of square pixels has a long tail to high multipoles. 

The choice of $\lambda$ to achieve rapid convergence 
is more of an art than a science.  We have found empirically that
it is sufficient to choose $\lambda$ so that at each 
iteration no $\epsilon\epsilon$
 parameter in the range $l_{\rm field} < l < l_{\rm pix}$ 
changes up or down by a factor of more than $3$.   
However, for bands that are noise dominated there are cases where
the maximum likelihood solution desires negative signal power.  For
this reason, we also impose a minimum $\lambda=0.01$ so that the power in
a band can be reduced below zero; we do require that the total power
in the signal and noise be positive by resetting negative values
to a small positive number at the start of each iteration.  
On top of these criteria, we halve $\lambda$
whenever $\delta p$'s from consecutive iterations cancel to $\approx 20\%$.
Since there is in principle no or little signal in the $\beta\beta$,
$\epsilon\beta$ and end $\epsilon\epsilon$ bands we stop iterating when
we reach better than 5\% convergence in all the remaining bands.  
The end bands should then be dropped when using the bandpowers
for cosmological constraints.

Finally it remains to estimate the errors on our maximum likelihood power
spectrum.  Ideally one would estimate the 2-point function of the parameters
$p_\alpha$ by Monte-Carlo integration
\begin{equation}
{\rm Cov}_{\alpha\beta}=
  \left< \delta p_\alpha \delta p_\beta \right> =
  \int d\vec{p}\quad \delta p_\alpha \delta p_\beta\ {\cal L}(p_\alpha)
\end{equation}
over the likelihood function.  However the likelihood evaluation is
sufficiently slow that this method impractical.  Instead we use the
Fisher matrix.
If the likelihood is sufficiently Gaussian (in the parameters) around the
maximum, one can estimate the covariance matrix from the curvature matrix,
\begin{eqnarray}
  {\cal F}_{\alpha \beta} &\equiv& - [\ln {\cal L}(p_\alpha)]_{, \alpha\beta}
  \\
	&=& 
  {\rm tr} [(d d^T - C)(C^{-1}C_{,\alpha}C^{-1}C_{,\beta}C^{-1} 
  \nonumber\\
	&&\quad - {1 \over 2} C^{-1} C_{,\alpha\beta} C^{-1})] +
	{1 \over 2} {\rm tr} (C^{-1} C_{,\alpha} C^{-1} C_{,\beta})\nonumber
\end{eqnarray}
or, assuming that that the maximum likelihood model is correct, from its
expectation value, the Fisher matrix,
\begin{eqnarray}
  {\rm Cov}_{\alpha\beta} \approx (F^{-1})_{\alpha\beta}\,.
\end{eqnarray}
Since the Fisher matrix is automatically calculated in the iteration
process, error estimates come at no additional computational expense.

\begin{figure*}[tbh]
\centerline{\epsfxsize=5.5truein\epsffile{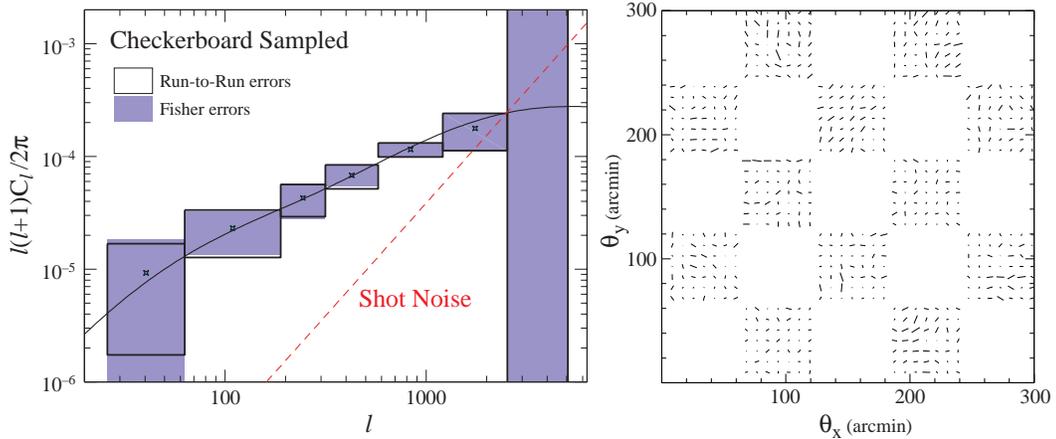}}
\caption{\footnotesize Checkerboard sampled Gaussian shear and 
recovered power $\epsilon\epsilon$ band powers from 100 realizations. 
The number and size of the pixels as well as the number and type of bands
are the same as in Fig.~\protect\ref{fig:sim}.  The mean power and errors
are well recovered in spite of the more complicated sampling.}
\label{fig:checker}
\end{figure*}

\section{Tests} \label{sec:tests}

There are several approximations that we have introduced in the above
algorithm, and it is of interest to ask how well the method works on
simulated data.  This also provides us an opportunity to demonstrate the
advantages of this method for realistic observational scenarios where
irregularly shaped fields complicate
the calculation of the large-angle correlation function or power spectrum.
These might arise from excising contaminated regions of the map or from
sparse sampling strategies.

\begin{table*}[bth]
\begin{center}
\caption{\label{tab:cov}}
{\sc Band Covariance\\}
\begin{tabular}{r|rrrrrrr}
\tablevspace{3pt}
\hline
band     & 37-90  & 90-270 & 270-450 & 450-669 & 669-1306 & 1306-2550 & 2550-5100 \\
\hline
37-90    & 1.00   & -0.14  & 0.23    & 0.12    & 0.13     & 0.07      & -0.08 \\
90-270   & (-0.15)&  1.00  & 0.08    & 0.20    & 0.17     & 0.05      &  0.02 \\
270-450  & (0.01) &(-0.11) & 1.00    & 0.37    & 0.43     & 0.25      & -0.15 \\
450-669  & (0.00) &( 0.00) & (-0.11) & 1.00    & 0.38     & 0.26      & -0.22 \\
669-1306 & (0.00) &(-0.01) & (0.00)  & (0.05)  & 1.00     & 0.50      & -0.38 \\
1306-2550& (0.00) &(-0.01) & (0.00)  & (0.03)  & (-0.31)  & 1.00      & -0.66 \\2550-5100& (0.00) &( 0.01) & (0.00)  & (-0.00) & (-0.40)  & (-0.68)   &  1.00 \\
\end{tabular}
\end{center}
\footnotesize NOTES.--- Covariance matrix of the $\epsilon\epsilon$ bands
recovered from $N$-body simulations. Upper: run-to-run covariance. Lower 
(parenthetical numbers): Fisher matrix estimates.  The Fisher matrix
underestimates the covariance in the intermediate regime where the signal is
mildly non-Gaussian and dominates the shot noise.
\end{table*}

\subsection{Gaussian Shear}

In Fig.~\ref{fig:sim}, we show a Gaussian realization of a 
shear power spectrum that corresponds to the $\Lambda$CDM cosmology 
with source galaxies at $z=1$ employed in \cite{WhiHu00} (2000). 
Gaussian distributed noise has been added to the 625 $\sim 8' \times 8'$ pixels 
corresponding to $\gamma_i^{\rm int}=0.4$
and $\bar n = 56$ gal/arcmin$^2$.   The theoretical signal and noise
power spectra are shown as lines in the left panel.
We choose 7 $\epsilon\epsilon$, 6 $\beta\beta$ and 6 $\epsilon\beta$ bands
according to the rules set down in the previous section.  The exact
$l$-ranges for the $\epsilon\epsilon$ bands are given in Table~\ref{tab:cov};
the other sets are similar save for the absence of the lowest $l$-band.
Note that $l_{\rm field}=106$ and $l_{\rm pix}=2550$. 
The recovered $\epsilon\epsilon$ power spectrum is 
shown in Fig.~\ref{fig:sim} (top left) with sampling error estimates per
realization obtained from averaging the Fisher variance estimates
and run-to-run scatter over 200 realizations.  Not shown are the
$\beta\beta$ and $\epsilon\beta$ bands where no statistically significant
power was recovered.
Notice that the two methods of estimating
the errors are in excellent agreement.  
The bands were chosen to be wide enough that their
correlation due to survey geometry is negligible.  
For these Gaussian
simulations, the covariance matrix of the bands reflects this fact
with negligible off-diagonal entries except for the end $\epsilon\epsilon$ bands
which have little intrinsic signal.

\subsection{$N$-body Shear}

In reality the shear field will be non-Gaussian due to the
non-linearity of the underlying density fields.
Because of the averaging effect of projection, 
the non-Gaussianity of the shear is much
milder than that of the density field. To test its effects on
the likelihood method, we use the simulations described in
\cite{WhiHu00} (2000).  We pixelize these simulations to the
same level as used for the Gaussian runs and add the same amount of
shot noise.
A sample shear field is shown in Fig.~\ref{fig:sim} (bottom right).
In the presence of this level of pixelization and noise, the 
non-Gaussianity of the N-body shear is in good agreement with that
estimated from higher resolution $N$-body simulations
(\cite{JaiSelWhi00} 2000; \cite{WhiHu00} 2000). 

In Fig.~\ref{fig:sim} (bottom left), we show the power spectrum
recovered from the simulated skies with error estimates as obtained
in the Gaussian simulations.   Non-Gaussianity does not bias the
power spectrum estimates.  Deviations in the lowest $l$ bin may
be attributed to finite box effects in the \cite{WhiHu00} (2000) simulations.

\begin{figure}[tbh]
\centerline{
\epsfxsize=3.25truein\epsffile{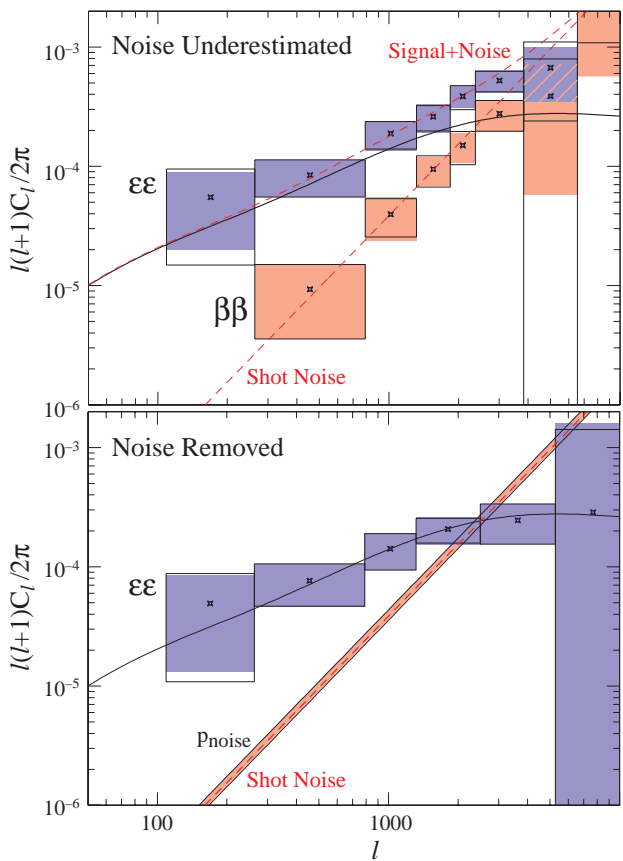}
}
\caption{\footnotesize Monitoring and removing excess noise. Top: Shot noise 
power underestimated
by a factor of 100.  The recovered 9
$\epsilon\epsilon$ and 8 $\beta\beta$ bandpowers
(end 2 bands of each off scale) 
show an excess that is equal in the two bands and rises as $l^2$
to the pixel scale $l_{\rm pix} \approx 7500$ (see text).  
Bottom: Shot noise underestimated but removed with an added white 
noise parameter $p_{\rm noise}$.
Both the 7 $\epsilon\epsilon$ bands (end band off scale)
{\it and} the excess white noise are well recovered.
Boxes and shading represent errors as in Fig.~\ref{fig:sim} and each
Gaussian realization of 60 utilizes 625 pixels. 
}
\label{fig:noise}
\end{figure}

Nevertheless the 
errors are slightly underestimated by the Fisher matrix 
in the intermediate regime 
where the intrinsic field is mildly non-Gaussian
and the removal of shot noise does not dominate the errors. 
The increased variance mainly arises from the covariance of the Fourier
modes within the bands induced by non-linear mode coupling effects in
the underlying density field.  Mode coupling also correlates the bands
themselves.  We show the covariance matrix Cov$_{\alpha\beta}/
({\rm Cov}_{\alpha\alpha} {\rm Cov}_{\beta\beta})^{1/2}$
of the $\epsilon\epsilon$ band powers in
Table~\ref{tab:cov} from the Fisher matrix and from the run-to-run covariance.
Again in the intermediate regime, the covariance of the bands is
underestimated by the Fisher matrix.

Although not a severe effect, this underestimation suggests that
in the absence of a sufficient number of fields from which the covariance
matrix may be extracted directly from the data, simulations and 
semi-analytic techniques (e.g. \cite{ScoZalHui99} 1999; \cite{CooHu00} 2000)
should be used to calibrate the covariance
matrix of band powers extracted from the data before using the measurements
to constrain cosmological parameters.

\subsection{Irregular sampling}

One of the advantages of likelihood based methods is that they automatically
account for any irregularity in the sampling or survey geometry, while
maintaining an optimal weighting of the data on each angular scale.
Sparse sampling techniques can be used to extend the dynamic range of power
spectrum estimates for the same amount of observing time (\cite{Kai98} 1998).
Irregular sampling can also be used to test the effect suspect regions of the
field might have on the results.

For illustration we test the method on a large field sampled in a
checker-board fashion with the same number and size of the pixels as above
(see Fig.~\ref{fig:checker} right panel).
Fig.~\ref{fig:checker} (left panel) shows the power spectrum recovered from
100 realizations along with error estimates as above.  Both the recovered
power and the errors agree well.

\subsection{Excess Noise}

Including bands for the $\beta\beta$ and $\epsilon\beta$ power spectra
is not strictly necessary if all sources of noise have been accounted
for in the noise covariance matrix.  However they do provide a means
of checking for any excess shot noise or systematic effects in the
data.  To illustrate this use, suppose that the initial estimate
of the shot noise contribution were low by a factor of $10$ 
(i.e.~by $100$ in power).  
As shown in Fig.~\ref{fig:noise} (top), the resulting power 
spectra in $\epsilon\epsilon$ and $\beta\beta$
show excess contributions which scale as $l^2$ in
bandpower and are equal in $\epsilon\epsilon$ and $\beta\beta$ above the
pixel scale.  Since a white noise spectrum is not well approximated
by a flat bandpower, errors in the pixel window function are exacerbated
leading to deviations near the pixel scale.

In the event that excess noise is detected and that its correlation
function or power spectrum may be parameterized, the likelihood technique 
can be easily modified to include and effectively marginalize these
noise parameters from the data itself.  The white noise case above
provides the simplest example where there is only one extra parameter
$p_{\rm noise}$, providing an addition to the signal covariance matrix
of the form
\begin{equation}
C^{\rm sig}_{(ij)(ab)} = {p_{\rm noise} \over N_i} \delta_{ij} \delta_{ab}\,.
\end{equation}
In Fig.~\ref{fig:noise} (bottom), we show the result of dropping the
$\beta\beta$ bands in favor of a white noise parameter.
Both the true signal and excess white noise are well recovered
by the technique.  The parameter $p_{\rm noise}$ is plotted 
as a power spectrum $C_l^{\rm noise} = p_{\rm noise} / \bar n$ with
error estimates from both the run-to-run scatter and the Fisher
matrix.  The covariance or degeneracy between the two is 
negligible since white noise has equal power in the two modes
whereas the true signal does not.  Excess noise from systematic errors will 
be more challenging to model, but the same principles and methods apply. 

\section{Conclusions} \label{sec:conclusions}

On large angular scales the shear field induced by weak gravitational lensing
of background galaxies by large-scale structure is close to Gaussian.  In
this regime the relevant information is encoded in the angular power spectrum,
$l(l+1)C_l/(2\pi)$.
We have suggested a technique, commonly used in CMB analysis, for determining
the angular power spectrum in this regime:
the use of ``band-powers'' extracted from the data by an iterated quadratic
estimator (\cite{BonJafKno98} 1998) of the maximum likelihood solution.
This technique has the advantage of automatically taking into account
irregular survey geometries and varying sampling densities.  It provides an
optimal estimate of the power spectrum which makes efficient use of all of
the data on the relevant angular scales.  We have tested the technique
against simulated Gaussian and realistically non-Gaussian data, regular
and irregularly sampled data, and with known and unknown amplitudes
of shot noise from the intrinsic ellipticities of galaxies.
In all cases, the mean band powers are recovered correctly.   

The technique introduced here is a result of one of many possible
cross-fertilizations of CMB and weak lensing research.  Indeed the
experience gained in measuring the shear power spectra from noisy
windowed data may feed back into the analogous problem for future
CMB polarization studies.
Techniques for handling large CMB data sets where the likelihood algorithm
used here becomes prohibitively expensive
(e.g.~\cite{WanHivGor} 2000; \cite{Szapudi} 2000)
will also be useful to lensing studies as the lensing fields become ever
larger.

\smallskip
{\it Acknowledgments:} 
W. Hu and M. White were supported by Sloan Fellowships.
M.~White was additionally supported by the US National Science Foundation.


\smallskip
\begin{thebibliography}{99}

\mybib{Bacon, Refregier \& Ellis}{BacRefEll00}
\aut{Bacon}{D} \aut{Refregier}{A} \laut{Ellis}{R}
\urefs{Detection of Weak Gravitational Lensing by Large-Scale Structure}
{\MNRAS}{submitted}{2000}{astro-ph/0003008}

\mybib{Blandford et al.}{BlaSauBraVil91}
\aut{Blandford}{R.D} \aut{Saust}{A.B} \aut{Brainerd}{T.G} \laut{Villumsen}{J.V} 
\refs{The distortion of distant galaxy images by large-scale structure}
{\MNRAS}{251}{600}{1991}{}

\mybib{Bond, Jaffe \& Knox}{BonJafKno98}
\aut{Bond}{J.R} \aut{Jaffe}{A.H} \laut{Knox}{L}
\refs{Estimating the Power Spectrum of the Cosmic Microwave Background}
{\PRD}{58}{083004}{1998}{astro-ph/9803272}

\bibitem[Bartelmann \& Schneider]{BarSch00}
   Bartelmann, M., \& Schneider, P. 2000, Phys. Rep., in press,
  astro-ph/9912508

\bibitem[Cooray \& Hu]{CooHu00}
   Cooray, A.R. \& Hu, W. 2000, in preparation

\mybib{Hu \& Tegmark}{HuTeg99}
\aut{Hu}{W}  \laut{Tegmark}{M}
\refs{Weak Lensing: Prospects for Cosmological Parameter Measurements}
{\ApJL}{514}{65}{1999}{astro-ph/9811168}

\mybib{Jain \& Seljak}{JaiSel97}
\aut{Jain}{B} \laut{Seljak}{U} 
\refs{Cosmological Model Predictions for Weak Lensing: Linear and Nonlinear Regimes}
{ApJ}{484}{560}{1997}{astro-ph/9611077}

\mybib{Jain et al.}{JaiSelWhi00}
\aut{Jain}{B} \aut{Seljak}{U} \laut{White}{S.D.M} 
\refs{Ray-tracing Simulations of Weak Lensing by Large-Scale Structure}
{ApJ}{530}{547}{2000}{astro-ph/9804238}

\mybib{Kaiser}{Kai92}
\aut{Kaiser}{N} 
\refs{Weak gravitational lensing of distant galaxies}
{ApJ}{388}{272}{1992}{}

\mybib{Kaiser}{Kai98}
\aut{Kaiser}{N} 
\refs{Weak Lensing and Cosmology}
{ApJ}{498}{26}{1998}{astro-ph/9610120}

\mybib{Kaiser, Wilson \& Luppino}{KaiWilLup00}
\aut{Kaiser}{N} \aut{Wilson}{G} \laut{Luppino}{G.A}
\urefs{Large-Scale Cosmic Shear Measurements}
{\ApJL}{submitted}{2000}{astro-ph/0003338}

\mybib{Mellier}{Mel99}
\aut{Mellier}{Y} 
\refs{Probing the universe with weak lensing}
{\ARAA}{37}{127}{1999}{astro-ph/9812172}

\mybib{Miralda-Escude}{Mir91}
\aut{Miralda-Escude}{J} 
\refs{The correlation function of galaxy ellipticities produced by gravitational lensing} 
{ApJ}{380}{1}{1991}{}

\mybib{Scoccimarro et al.}{ScoZalHui99}
\aut{Scoccimarro}{R} \aut{Zaldarriaga}{M} \laut{Hui}{L} 
\refs{Power Spectrum Correlations Induced by Nonlinear Clustering}
{ApJ}{527}{1}{1999}{astro-ph/9901099}

\mybib{Seljak}{Sel98}
\aut{Seljak}{U} 
\refs{Weak Lensing Reconstruction and Power Spectrum Estimation: Minimum Variance Methods}
{ApJ}{506}{64}{1998}{astro-ph/9711124}

\mybib{Szapudi et al.}{Szapudi}
\aut{Szapudi}{I} et al.
\urefs{Fast CMB Analyses via Correlation Functions}{\ApJ}
{submitted}{2000}{astro-ph/0010256}

\mybib{van Waerbeke et al.}{vanetal00}
\aut{van Waerbeke}{L} et al.
\refs{Detection of correlated galaxy ellipticities from CFHT data:
        first evidence for gravitational lensing by large-scale structures}
{\AsAs}{358}{30}{2000}{astro-ph/0002500}

\mybib{Wandelt, Hivon \& Gorski}{WanHivGor}
\aut{Wandelt}{B} \aut{Hivon}{E} \laut{Gorski}{K}
\urefs{The Pseudo-$C_l$ method}{\PRD}{submitted}{2000}{astro-ph/0008111}

\mybib{White \& Hu}{WhiHu00}
\aut{White}{M} \laut{Hu}{W}
\refs{A New Algorithm for Computing Statistics of Weak Lensing by
Large-Scale Structure}{\ApJ}{537}{1}{2000}{astro-ph/9909165}

\mybib{Wittman et al.}{Witetal00}
\aut{Wittman}{D.M} \aut{Tyson}{J.A} \aut{Kirkman}{D}
 \aut{Dell'Antonio}{I} \laut{Bernstein}{G}
\refs{Detection of weak gravitational lensing distortions of distant
        galaxies by cosmic dark matter at large scales}
{Nature}{405}{143}{2000}{astro-ph/0003014}

\end{thebibliography}
\end{document}